# Integrating DRAM Power-Down Modes in gem5 and Quantifying their Impact

Radhika Jagtap
Arm Ltd.
Cambridge, U.K.
radhika.jagtap@arm.com

Matthias Jung
Fraunhofer IESE
Kaiserslautern, Germany
matthias.jung@iese.fraunhofer.de

Wendy Elsasser
Arm Inc.
Austin, TX, U.S.A.
wendy.elsasser@arm.com

Christian Weis
University of Kaiserslautern
Kaiserslautern, Germany
weis@eit.uni-kl.de

Andreas Hansson
Arm Ltd.
Cambridge, U.K.
andreas.hansson@arm.com

Norbert Wehn
University of Kaiserslautern
Kaiserslautern, Germany
wehn@eit.uni-kl.de

## ABSTRACT

Across applications, DRAM is a significant contributor to the overall system power, with the DRAM access energy per bit up to three orders of magnitude higher compared to on-chip memory accesses. To improve the power efficiency, DRAM technology incorporates multiple power-down modes, each with different trade-offs between achievable power savings and performance impact due to entry and exit delay requirements. Accurate modeling of these low power modes and entry and exit control is crucial to analyze the trade-offs across controller configurations and workloads with varied memory access characteristics. To address this, we integrate the power-down modes into the DRAM controller model in the open-source simulator gem5. This is the first publicly available *full-system* simulator with DRAM power-down modes, providing the research community a tool for DRAM power analysis for a breadth of use cases. We validate the power-down functionality with sweep tests, which trigger defined memory access characteristics. We further evaluate the model with real HPC workloads, illustrating the value of integrating low power functionality into a full system simulator.

## 1 INTRODUCTION

System designers usually focus on optimizing for performance, which means that they *"make the common case fast"*. But the common case for real systems like a smart phone or web server is very often *"doing nothing"*. However, doing nothing could consume a significant amount of standby power. Therefore, we need to design for the common case from a power perspective, which means that we have to *"do nothing well"* [1]. For the design of DRAM subsystems this translates to efficient power-down strategies, which are the only way to reduce the power consumption when there is no memory activity.

The usage profiles of typical smartphone users show a burst-like behavior [11]. Users expect to resume their applications in the state last used. When the phone is not in use, applications therefore store their data in the memory. Both, the short bursts of intensive smart-phone usage and the standby or idle phases require power optimization. For production servers and data centers, analysis has shown that servers are idle over 60% of the time. Further, DRAM consumes 20% to 40% of the total system power [36] and enabling DRAM power-down modes (i.e. power-down and self-refresh) is a must for effective cross-layer energy efficiency solutions [36]. For example, self-refresh mode saves a significant amount of power and transitions into and out of self-refresh can be completed in less than a microsecond [36].

In summary, the utilization of the available DRAM bandwidth and the use of power-down modes are major contributions to high energy efficiency of DRAM subsystems and in turn of complete integrated systems. In existing publicly available DRAM simulators, low power modes are either missing or only a subset of them are modelled.

In this paper, we describe the low power DRAM functionality we contributed to the open-source simulator gem5 [3]. To the best of our knowledge, the code we released makes gem5 the first publicly available full-system simulator featuring power-down modes. gem5 also has a well correlated DRAM power tool DRAMPower [6] integrated within it making it a complete framework for energy efficiency studies. We validate the power-down functionality by a thorough characterization involving sweep testing. Additionally, we evaluate the model using realistic workloads from the *High Performance Computing* (HPC) domain. Our contribution enables energy efficiency studies for a broad range of systems spanning smartphones, servers and HPC which would be tedious at best if not impossible due to proprietary nature of available tools. In summary, our contributions are as follows.

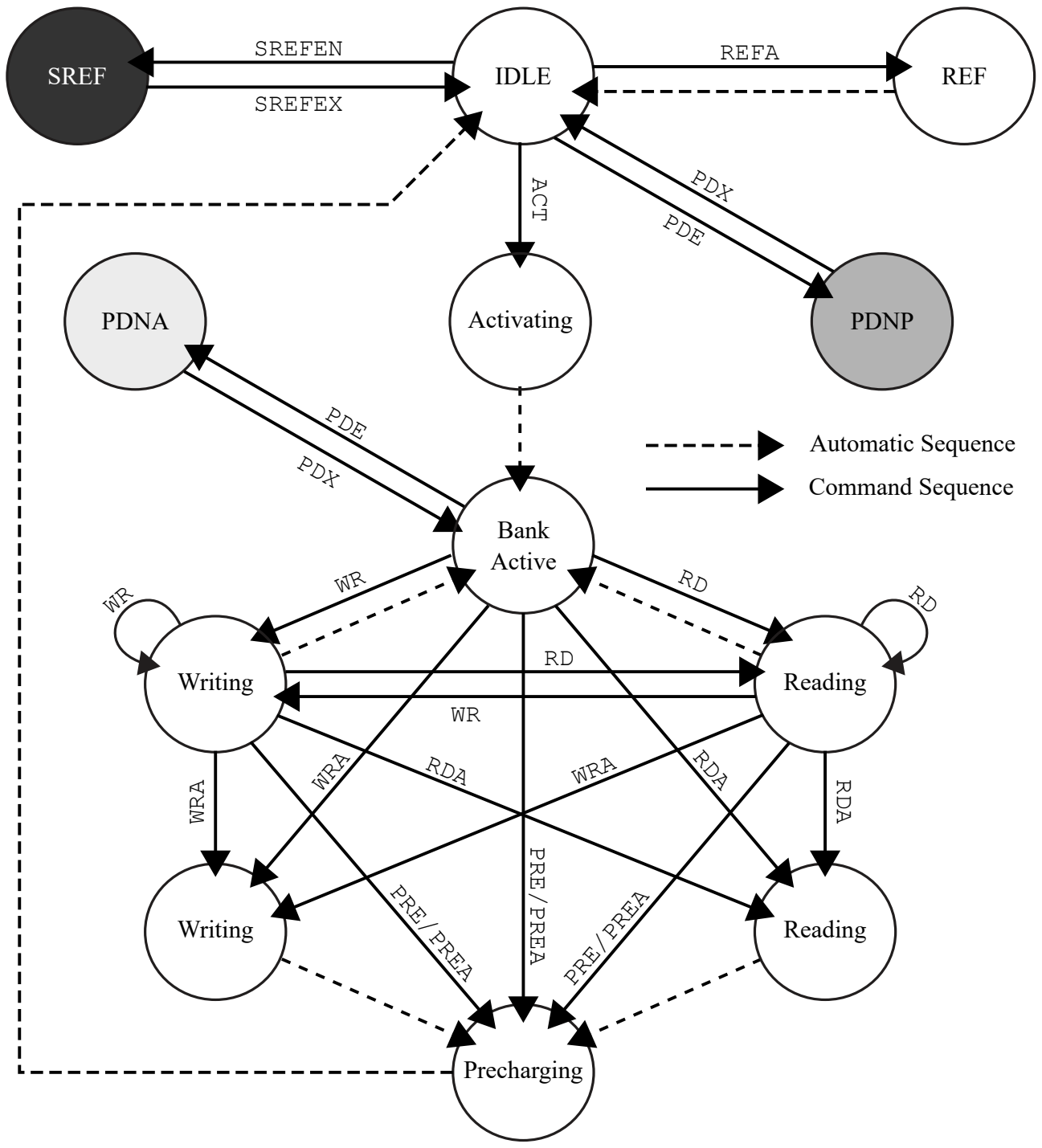

**Figure 1: State diagram of DRAM commands according to JEDEC [33], power-down modes highlighted in grey [23]**

(1) release of DRAM power-down modes in gem5
(2) model validation using low-level sweep testing
(3) model evaluation using realistic workloads

The remainder of this paper is organised as follows. The background and related work is covered in Sections 2 and 6, respectively. We describe our power state model in Section 3 and its behavioural validation in Section 4. Finally we evaluate our model using real workloads in Section 5 and conclude in Section 7.

## 2 BACKGROUND

In today's *Systems on Chip* (SoCs) DRAM is a major contributor to the total power consumption. For instance, the authors of [4] show in a power breakdown of a recent smartphone that DRAM contributes around 17% to the total system power. Moreover, there are applications, such as the GreenWave computing platform [29], in which 49% of the total power consumption has to be attributed to DRAMs, and even 80% for a system that imitates the human cortex based on an *Application Specific Integrated Circuit* (ASIC), as shown in [13]. In fact, the energy consumed per bit for accessing off-chip DRAM is two to three orders of magnitude higher than the energy required for on-chip memory accesses [27]. This is due to complex and power hungry I/O transceiver circuits that have to deal with the electrical characteristics of the high-speed interconnections (transmission lines) between the chips.

Furthermore, DRAMs must be refreshed regularly due to their charge based bit storage property (capacitor). This DRAM refresh operation must be issued periodically and causes both performance degradation, with respect to bandwidth and latency, and increased energy consumption. For example, if we consider a computer with 4TB of DRAM, turning the DRAM on will immediately draw $\approx$ 290 W of standby power[1]. However if the self-refresh power-down mode is used the standby power is reduced to $\approx 200\,\text{W}$[2].

In summary, energy consumption of DRAM memory sub-systems has become a significant concern for a broad range of use-cases. By utilizing the available DRAM bandwidth efficiently and by putting the DRAM device in power-down modes when there are no memory requests, we can achieve higher overall system energy efficiency.

DRAMs are organized in a three-dimensional fashion of banks, rows and columns. A DRAM device has usually eight (DDR3) or 16 (DDR4) banks, which can be used concurrently (*bank parallelism*). However, there are some constraints due to the shared data command/address bus. Each bank consist of e.g. $2^{12}$ to $2^{18}$ rows, whereas each row can store e.g. 512 B to 2 KB of data. The task of the DRAM controller is to translate incoming read and write requests to a sequence of DRAM commands. The controller must issue commands based on the state of the device and honor timing constraints for the specific DRAM standard defined by JEDEC.

Figure 1 shows a simplified state diagram including the states and the commands to transition between states as per JEDEC [33]. To access data in a row of a certain bank, the *activate* command (ACT) must be issued by the controller before any column access, i.e. *read* (RD) or *write* command (WR) can be executed. The ACT command opens an entire row of the memory array, which is transferred into the bank's *row buffer*[3]. It acts like a small cache that stores the most recently accessed row of the bank. If a memory access targets the same row as the currently cached row in the buffer (called *row hit*), it results in a low latency and low energy memory access. Whereas, if a memory access targets a different row as the current row in the buffer (called *row miss*), it results in higher latency and energy consumption. If a certain row in a bank is active it must be precharged (PRE) before another row can be activated. In addition to RD and WR commands, there exist read and write commands with an integrated auto-precharge, i.e. RDA and WRA. If auto-precharge is selected, the row being accessed will be precharged at the end of the read or write access without a PRE command (shown in Figure 1 by the dotted line arrows). In that context the term *page policy* is used.

---

[1] The standby power can be calculated as:

$$P = 512 \cdot \frac{t_{RFC} \cdot (V_{DD} \cdot I_{DD5B} + V_{PP} \cdot I_{PP5B}) + V_{DD} \cdot (t_{REFI} - t_{RFC}) \cdot I_{DD2N}}{t_{REFI}}.$$

For 512 DRAM DIMMs with size of 8GB each (e.g. Micron MTA8ATF1G64AZ [34]) and the following parameters: $I_{DD5B} = 1800\,\text{mA}$, $I_{PP5B} = 240\,\text{mA}$, $I_{DD2N} = 400\,\text{mA}$, $t_{RFC} = 350\,\text{ns}$, $t_{REFI} = 7.8\,\mu\text{s}$, $V_{DD} = 1.2\,\text{V}$, $V_{PP} = 2.5\,\text{V}$, $f = 1200\,\text{MHz}$ the resulting power is $P = 289.24\,\text{W}$.

[2] The self-refresh power can be calculated to a first approximation as:

$$P = 512 \cdot (V_{DD} \cdot I_{DD6} + V_{PP} \cdot I_{PP6}) = 198\,\text{W}.$$

where $I_{DD6} = 240\,\text{mA}$ and $I_{PP6} = 40\,\text{mA}$ for the example DIMM [34].

[3] The row buffer is a model, which abstracts the real physical DRAM architecture (primary and secondary sense amplifiers). For further details on internal DRAM architecture we refer to [18, 21].

There are two basic page policies called the *Open Page Policy* (OPP) and the *Close Page Policy* (CPP). The OPP keeps the current row active after a RD or WR, whereas the CPP precharges the row automatically using the RDA and WRA commands. The CPP is often used for server applications (e.g. webserver), where the accessed DRAM addresses are typically uniformly random. The OPP is used in desktop PCs and mobile devices where row hits are more likely due to a higher data locality. The improved *Adaptive Page Policy* (APP) keeps the row open if there are already queued accesses to the open row [14] and can be used in combination with scheduling.

As already mentioned, a DRAM cell must usually be refreshed every 64 ms to retain the data stored in it. Modern DRAMs are equipped with an *Auto-Refresh* (REFA) command to perform this operation. Besides the normal active mode operations presented above, a DRAM is capable of entering power-down modes to save energy by setting the clock-enable signal cke to low. There exist three major power-down modes called *Precharge Power-Down* (*PDNP*), *Active Power-Down* (PDNA) and *Self-Refresh* (SREF). Thus, a device can be in one of five states defined below and shown in Figure 1 [23].

- **Active:** At minimum one bank is active, no power-down (cke=1), no internal refresh (the DRAM controller has to schedule refresh commands).
- **IDLE:** All banks are closed and precharged, no power-down (cke=1), no internal refresh. The DRAM changes the state from *Active* to *IDLE* by issuing a precharge command (PRE).
- **Precharge Power-Down (PDNP):** All banks are closed and precharged (in *IDLE* state, cke=0) and no internal refresh.
- **Active Power-Down (PDNA):** At minimum one bank is active (in *Active* state, cke=0) and no internal refresh.
- **Self-Refresh (SREF):** All banks are precharged and closed, the DRAM internal self-timed refresh is triggered (cke=0).

The power saving potential depends on the duration of each power-down mode. Power-down modes could reduce performance because of their non-zero *exit times*. Therefore, the power-down functionality in the controller must identify the optimal point to enter a power-down state. The conditions for transitioning to a power-down state include an empty request queue and no pending events. In some controllers, after such conditions are met there is a timeout counter to delay the entry into a power-down mode. This is to ensure that the compute system is idling so as to increase chances of staying in power-down mode for long enough to compensate performance degradation by energy savings. As self-refresh is a deeper power-down mode than precharge power-down and activate power-down, it can take several clock cycles (DDR4 = 408, DDR3 = 512, Wide I/O = 20) to exit from SREF.

## 3 CONTROLLER POWER STATE MODEL

While the power state machine, the commands and the timings are dictated by JEDEC standards and device vendors, controller designers implement their own power-state entry and exit control. In this work we build this control logic on top of the DRAM controller model in gem5. The controller schedules a sequence to transition to power-down mode if *both conditions are true*:

(1) there are no requests in the queue
(2) there are no pending events

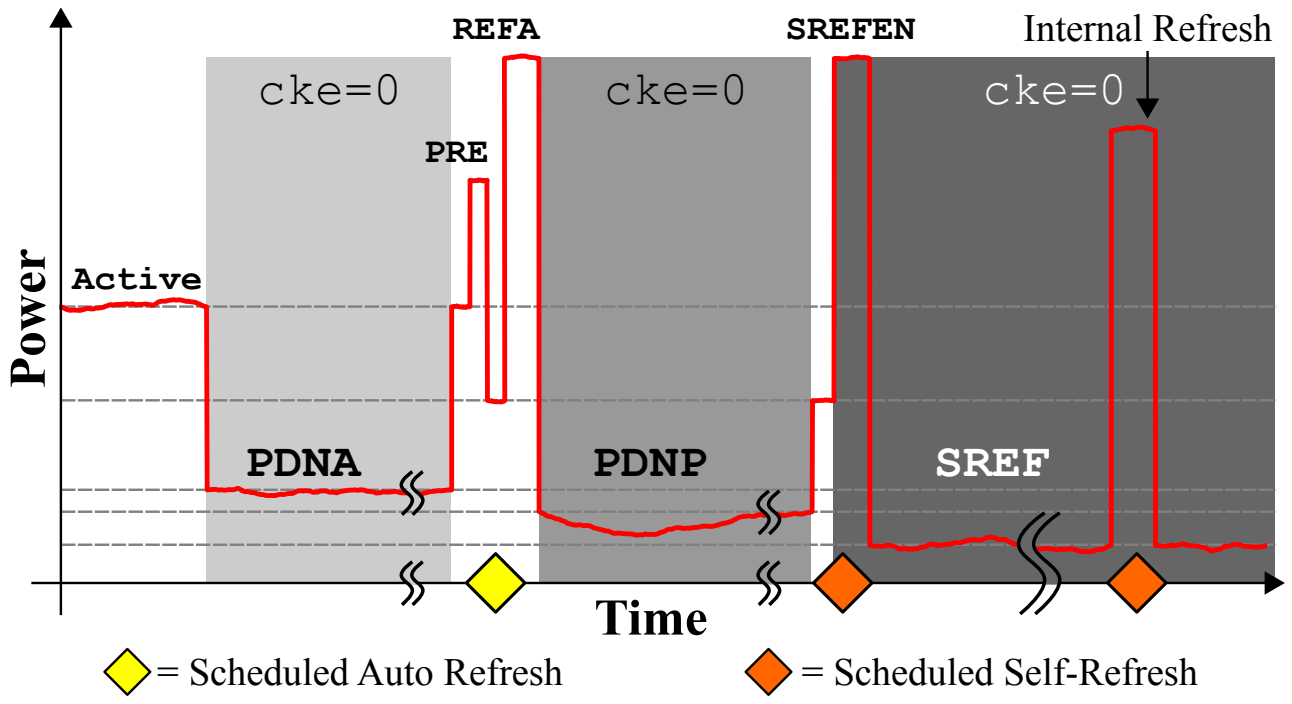

**Figure 2: Staggered power-down policy [26]**

Examples of pending events are an ongoing refresh operation and an impending precharge. Our model is event driven as against cycle callable. We check the aforementioned conditions required to transition to a power-down state on events such as start of refresh, end of refresh, respond with data and end of precharge.

As shown in Figure 1, the device can enter PDNA from an Active state (at least 1 bank has a row open). If the bank is closed and the device is in IDLE state, and the aforementioned conditions to enter a power-down state are met then the device can enter PDNP. From PDNP, on a refresh event the controller has the opportunity to check the conditions and put the device into the deeper power-down mode of self-refresh.

A non-optimized highly opportunistic self-refresh entry policy results in an increased average power, which should be avoided. This higher power consumption can be explained by the fact that each self-refresh entry provokes a refresh at the beginning. This increase in DRAM energy consumption was already measured and investigated by Schmidt et al. in [41]. They presented the overestimation of power savings in the Micron's power calculator [37] when using the DRAM self-refresh mode intensively.

Figure 2 shows the *Staggered* power-down strategy presented by Jung et al. [26]. After a read or write access the DRAM stays in Active mode (at least one bank active), in this strategy. However, if no new transaction is scheduled, the DRAM controller immediately sets cke to "0" and the DRAM enters active power-down mode (PDNA). After a certain time a refresh must be issued to the DRAM. Therefore, the controller has to wake up the DRAM with a power-down exit command (PDX), send a precharge-all command (PREA) and finally the refresh command (REFA). The controller switches to precharge power-down mode (PDNP), due to the previously executed PREA command. If there is still no new read or write request and the next refresh should be triggered (e.g. 7.8$\mu s$ later), the controller schedules a self-refresh entry (SREFEN) instead of a normal refresh command. This sequence is the key to the additional savings with staggered power-down policy. The controller uses the refresh commands as a trigger to enforce the state changes, PDNA⟶PDNP⟶SREF, to minimize the energy consumption of the DRAM. With this method, unnecessary SREF entries will be avoided, and the hardware timeout counters, as used in state-of-the-art controllers, are not required.

**Table 1: Power and timing parameters based on the Micron DDR4-2400 8 Gbit datasheet[35]**

| Current | Description | Value in mA |
|---|---|---|
| $I_{DD0}$ | Active precharge current | 43 |
| $I_{PP0}$ | Active precharge current for $V_{PP}$ | 3 |
| $I_{DD2N}$ | Precharge standby current | 34 |
| $I_{DD3N}$ | Active standby current | 38 |
| $I_{PP3N}$ | Active standby current for $V_{PP}$ | 3 |
| $I_{DD2P}$ | Precharge powerdown fast | 25 |
| $I_{DD3P}$ | Active powerdown fast | 32 |
| $I_{DD5}$ | Refresh current | 250 |
| $I_{DD6}$ | Self-refresh current | 30[4] |
| $I_{DD4R}$ | READ current | 110 |
| $I_{DD4W}$ | WRITE current | 103 |
| **Voltage** | **Description** | **Value in V** |
| $V_{DD}$ | Main power supply voltage | 1.2 |
| $V_{PP}$ | Activation power supply voltage | 2.5 |
| **Timing** | **Description** | **Value in ns** |
| $t_{CK}$ | DRAM clock period | 0.833 |
| $t_{CCD}$ | Column to column delay | 3.332 |
| $t_{RP}$ | Precharge to subsequent activate | 14.160 |
| $t_{RAS}$ | Min. time between an activate and precharge to the same row | 32.000 |

## 4 BEHAVIOURAL VALIDATION

We validate the behaviour of the DRAM controller using synthetic traffic as this provides a fine-grained control of the memory access pattern.

To trigger power-down mode transitions, the main parameter we need to manipulate is the inter-request or *Inter-Transaction Time* (ITT). As we increase ITT, the bus utilization and the occupancy of the controller's request queue reduces. At events such as the refresh event, if there are no outstanding read or write requests in the controller's queue, there is an opportunity to transition into a low power mode after refresh is complete [26]. In this way, as ITT increases the time spent in power-down modes increases.

For 100% bus utilization, the ITT must be equal to the time to the column to column delay $t_{CCD}$. Therefore we set the minimum value of ITT $ITT_{min}$ to $t_{CCD}$. After an ACT command is serviced and the row is precharged, the time between this ACT command and the earliest opportunity to transition to a power-down state is: $t_{PDE} = t_{RAS} + t_{RP} + t_{CK}$. We use this time to power-down mode entry, $t_{PDE}$, as a time unit to tune the value of ITT. During a phase, the traffic generator selects a random value between $ITT_{min}$ and a specified maximum value of ITT, termed $ITT_{max}$, as the delay between back-to-back requests. For each phase, we also configure a target bank utilization. For example by setting target utilization to 8/16, request addresses are generated for banks 0-7 only. Finally, we set the number of bytes sequentially accessed ($N_{SeqBytes}$) by a traffic generator request. In summary, we sweep a few parameters to tune the likelihood to entering power-down mode and the parameter combinations result in many traffic generation phases.

[4]In Table 1, IDD6 is chosen for the temperature range 0-85 C

**Table 2: Traffic generation parameters**

| **Memory configurations - total 4 simulations** | |
|---|---|
| no. of ranks | 1, 2 |
| page policy | open adaptive, closed adaptive |
| **Traffic parameters in each memory configuration** | |
| request type | read (always) |
| request size | 64 B |
| address range | 0 to 256 MB |
| address map | RoRaBaCoCh |
| $ITT_{min}$ | $t_{CCD}$ |
| $ITT_{max}$ | $t_{PDE}$, 20x $t_{PDE}$, 100x $t_{PDE}$ |
| $N_{SeqBytes}$ | 64, 256, 512 |
| bank utilisation | 1/16, 8/16, 16/16 |

### 4.1 Experimental Setup

The system we simulate in gem5 consists of a single traffic generator and a system bus that connects it to a DRAM controller. The simulated DDR4 DIMM (×64) consists of 16 chips each having a 4-bit interface (×4), with timings based on the Micron DDR4-2400 8 Gbit datasheet (Micron MT40A2G4) [35] connected to the controller via a single DDR4-2400 x64 channel. The key current values, based on the same datasheet, are shown in Table 1. These are used by the DRAMPower tool [6] integrated in gem5 to calculate the energy components. The commands and timestamps are input to DRAMPower at run time during simulation. The DRAM sub-system has 16 banks. The buffer containing the incoming requests for all banks is split into a read and write queue. Requests are re-ordered by the controller. The scheduling policy for the controller is *First Ready - First Come First Served* (FR-FCFS) [39]. With regards to the page policy, we explore two specializations of the APP described in Section 2, which are as follows:

- **Closed adaptive:** the page of a specific bank is closed if there are no requests to this bank or if there are accesses for this bank which are not row hits (i.e. there are row misses)
- **Open adaptive:** the page of a specific bank is closed if there are no row hits (i.e. there are row misses) but is kept open if there are no requests to this bank

More information about the controller in gem5 is provided in [14].

We set the number of ranks $N_{Ranks}$ to one and two and switch between the open adaptive and the closed adaptive page policies. In this way, we explore four memory configurations. We stimulate each memory configuration with our low-power traffic generation setup described earlier. The details of the traffic and the memory configurations are shown in Table 2. We set $ITT_{max}$ to generate the following profiles

- **Very dense:** $ITT_{max} = t_{PDE}$
- **Dense:** $ITT_{max} = 20 \cdot t_{PDE}$
- **Sparse:** $ITT_{max} = 100 \cdot t_{PDE}$

Within each density profile, we assign three values each for $N_{SeqBytes}$ and bank utilization, which gives us a total 27 traffic generation phases per memory configuration. Each phase is 250 usec long. Next, we discuss the results of the simulation.

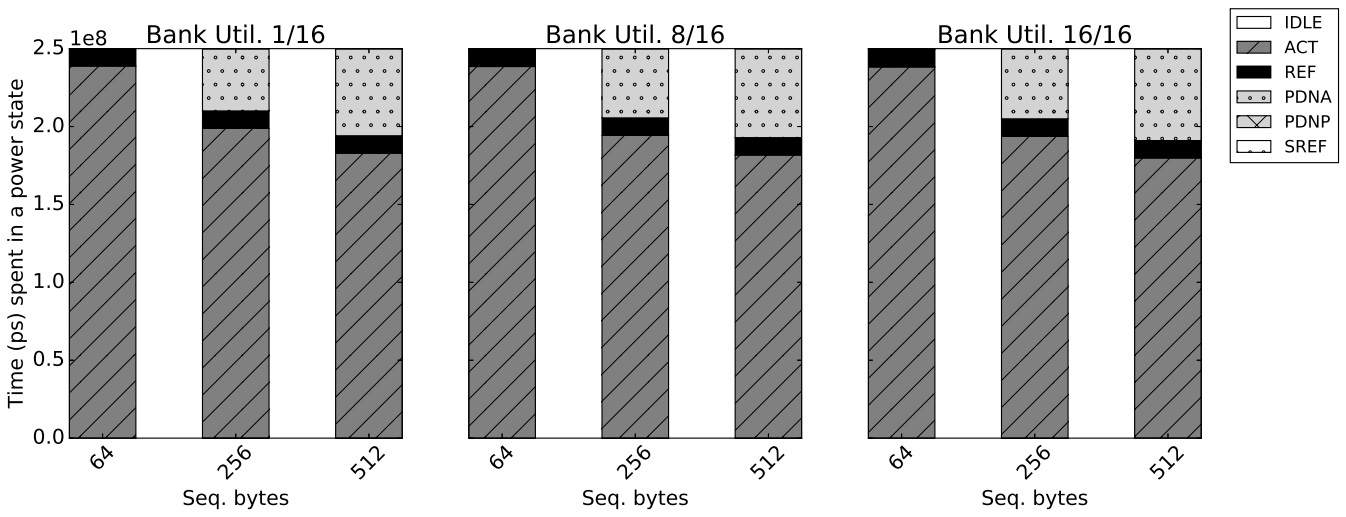

**(a) Time spent in power states**

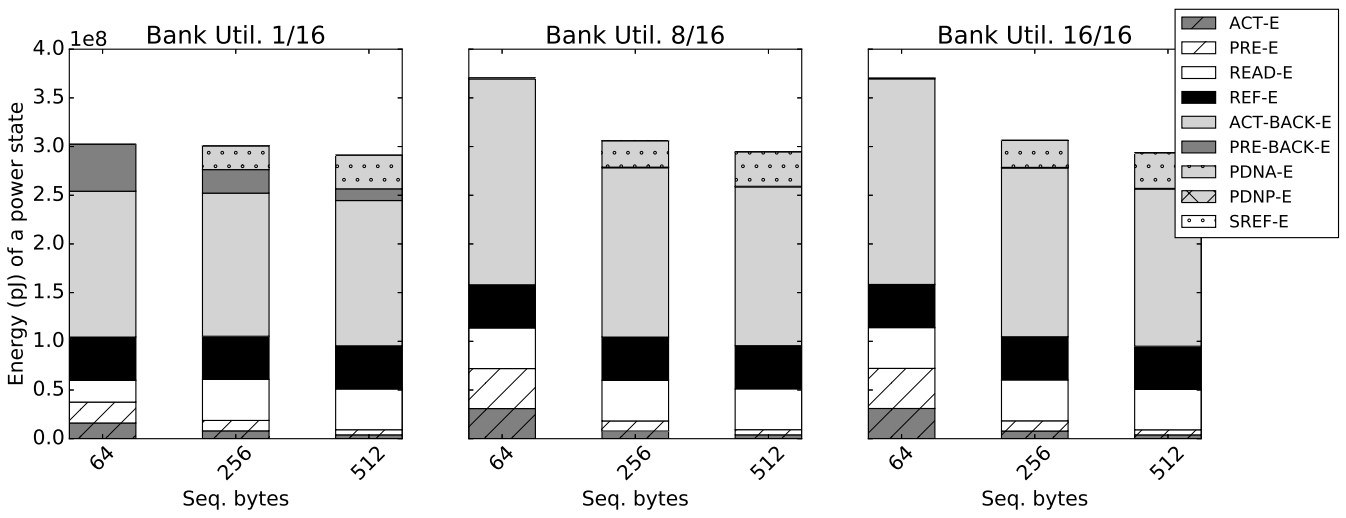

**(b) Energy consumed by power states**

**Figure 3: $N_{Ranks}$ = 1, open adaptive policy, very dense traffic**

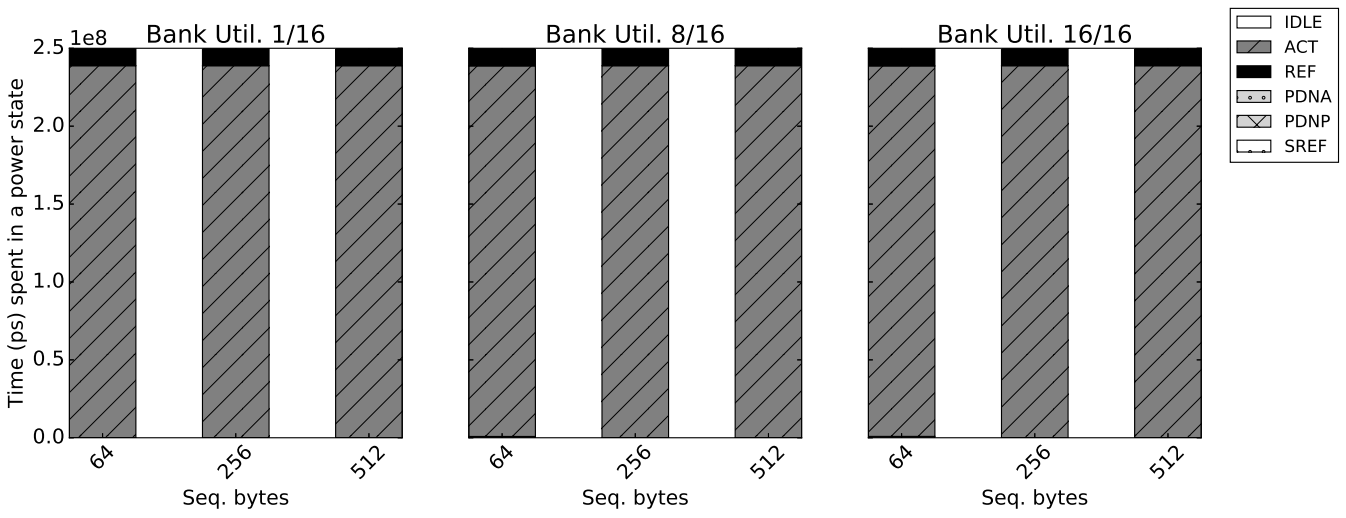

**(a) Time spent in power states**

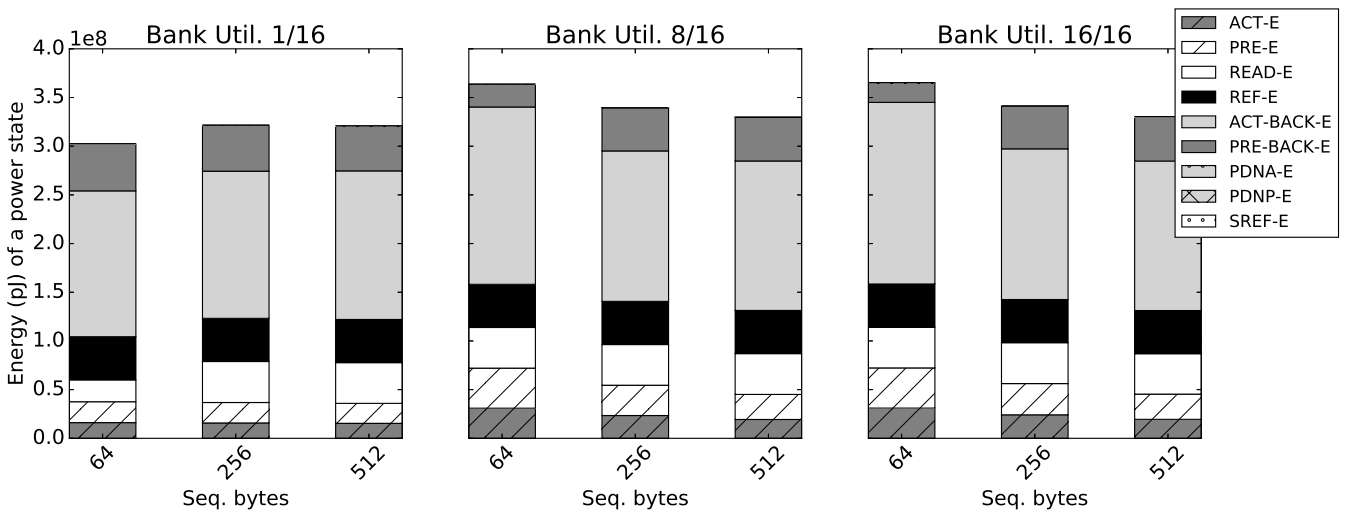

**(b) Energy consumed by power states**

**Figure 4: $N_{Ranks}$ = 1, closed adaptive policy, very dense traffic**

## 4.2 Results for very dense traffic

Figures 3a and 3b show the stack plots for the time spent and the energy consumed for memory configuration comprising one rank and the open adaptive page policy. When the number of sequential bytes accessed $N_{SeqBytes}$ increases, the time spent in PDNA mode (light grey with dots) increases. The request size of the system is 64B. When the $N_{SeqBytes}$ is set to 64, every new request is to a random address causing maximum row misses in the DRAM. On the other hand, when $N_{SeqBytes}$ is set to 256 and 512, the row hit rate is 3/4 and 7/8 respectively thus resulting in quicker servicing of the requests and in turn lower queue occupancies. For the 256 and 512 settings the controller's conditions to transition to power-mode are satisfied and given the open adaptive page policy we see that the DRAM goes into PDNA mode.

Figure 3b shows that the active background energy ACT-BACK-E (in light grey) is the biggest component. For bank utilization 8/16 and 16/16, the energy consumed reduces as $N_{SeqBytes}$ goes from 64B to 512B due to PDNA mode. This trend in energy consumption does not hold for bank utilization 1/16 due to cold start. In the first traffic phase, i.e. bank utilization 1/16 and $N_{SeqBytes}$ 64B, only half as many requests are issued compared to the remaining 8 phases. The controller queue occupancy is nearly always 100% in the first phase and traffic generator faces severe back pressure. This explains the lower energy consumption for the first phase. The refresh energy is constant as the number of refresh commands issued in each phase is constant.

Figures 4a and 4b show the time and energy stack plots for the closed adaptive page policy. As $N_{SeqBytes}$ goes from 64B to 512B, there is negligible time spent in power down mode in contrast to Figure 3a. The difference is that in closed adaptive policy, a precharge event is scheduled to close the page if there are no row hits in the queue. The controller prioritizes this precharge event over a PDNA transition. The number of times the controller detected no queued requests and one outstanding event was about 50 times that of the open adaptive page policy. Of these, in 90% of the cases the outstanding event was a precharge event. As expected, comparing the energy data with the open adaptive policy, active power-down energy (PDNA-E in light grey with dots) is traded for precharge and background precharge energy (PRE-E with hatch pattern and PRE-BACK-E in dark grey), thus resulting in higher total energy for closed adaptive policy.

## 4.3 Results for dense and sparse traffic

We discuss the data for two ranks instead of one as it shows more variation across the page policies and more time is spent in SREF compared to the simulation with one rank. The time spent in different states and the energy consumed are shown in Figures 5 and 6. The total energy consumed for dense and sparse traffic scenarios is in the range of 1.7 to 2.0 (Figure 6) which is significantly smaller than the range 2.5 to 3.0 pJ observed for very dense traffic for two ranks (not shown in figures). In case of both open adaptive and closed adaptive page policies, the total energy savings achieved when going from dense to sparse traffic are in the range of 5% to 15%.

For sparse traffic for the open adaptive page policy (Figure 6a), when the energy for self-refresh increases (SREF-E in white with dots) the energy for refresh reduces (REF-E in black). This validates the staggered power-down strategy implemented in the power state model, which is described in Section 3. For example, for bank utilization 8/16, self-refresh energy increases and refresh energy reduces as $N_{SeqBytes}$ increases from 64B to 512B. The activate background energy (ACT-BACK-E in light grey) is also reduced. For

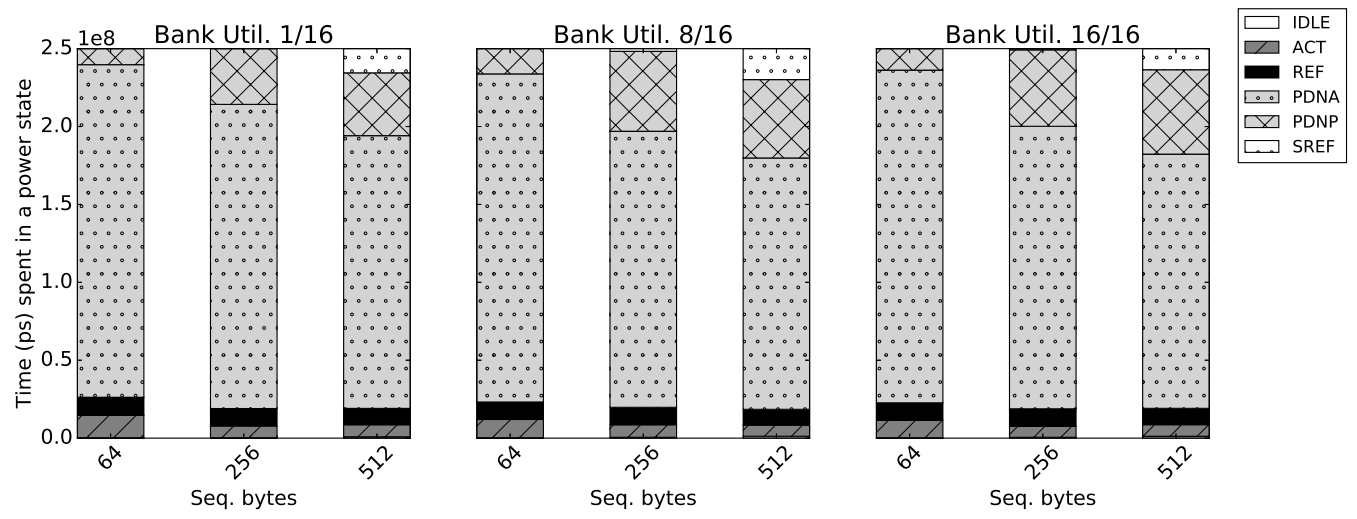

(a) Open adaptive policy, dense traffic

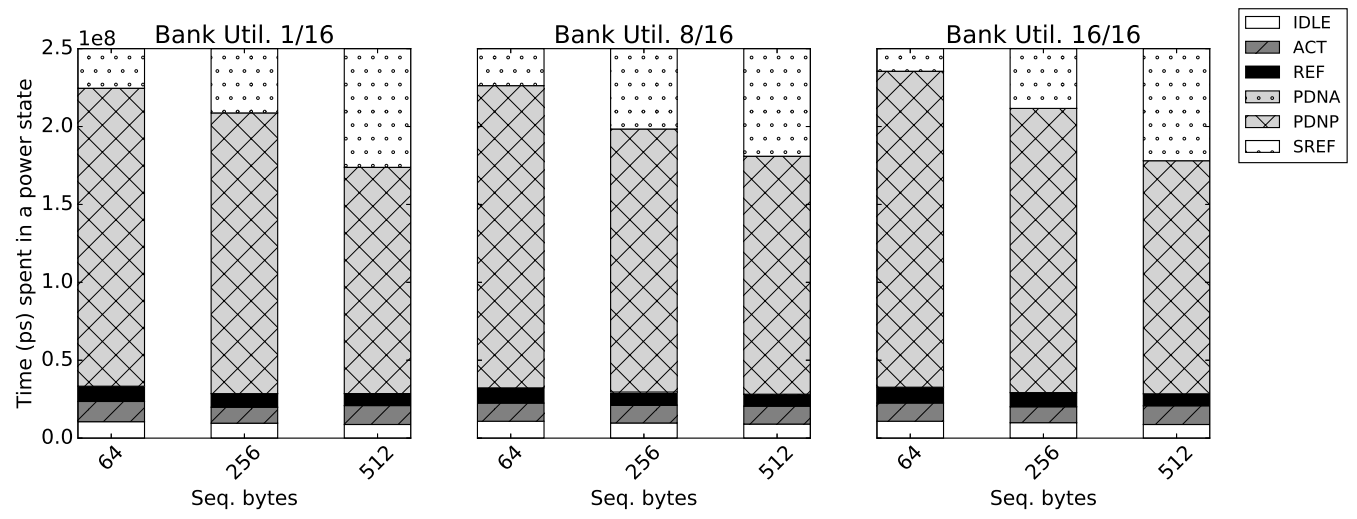

(b) Closed adaptive policy, dense traffic

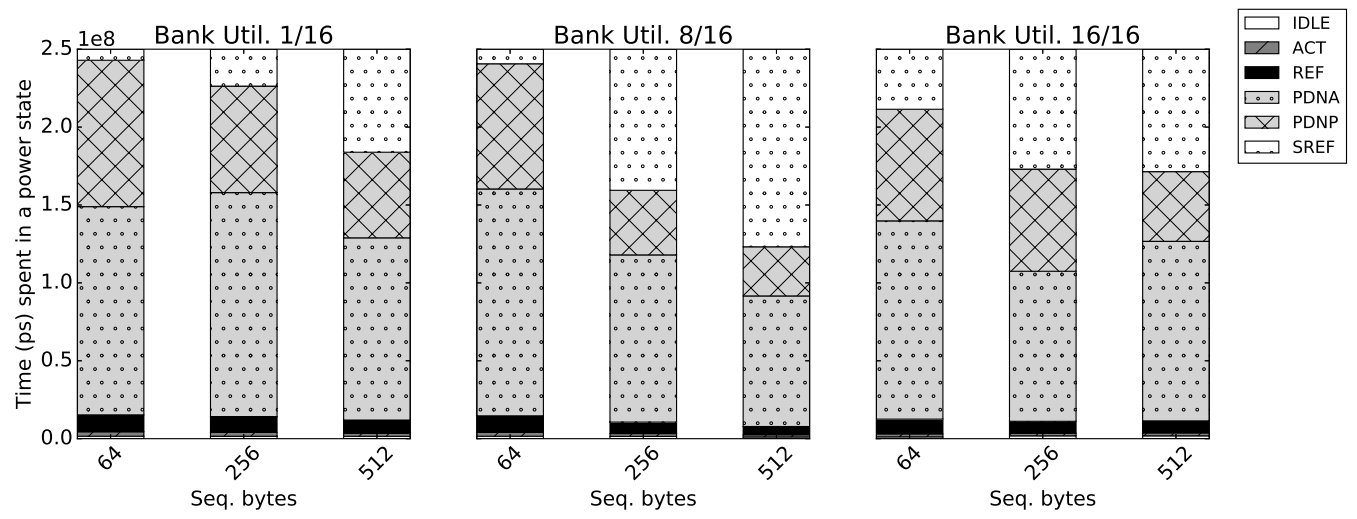

(c) Open adaptive policy, sparse traffic

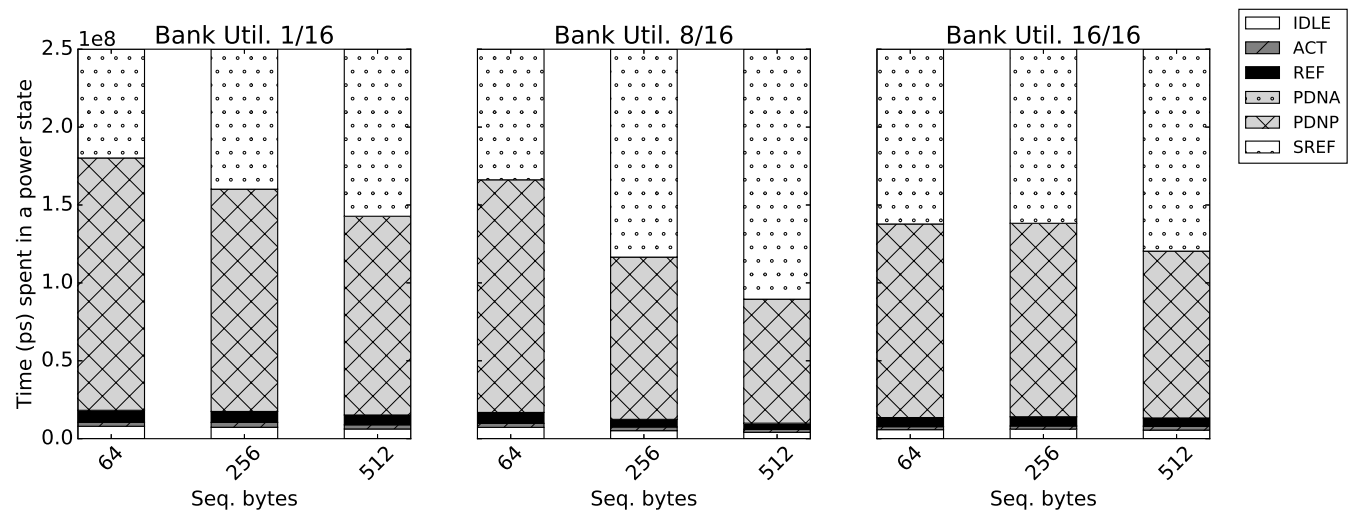

(d) Closed adaptive policy, sparse traffic

**Figure 5: Time spent in power states for $N_{Ranks}$ = 2**

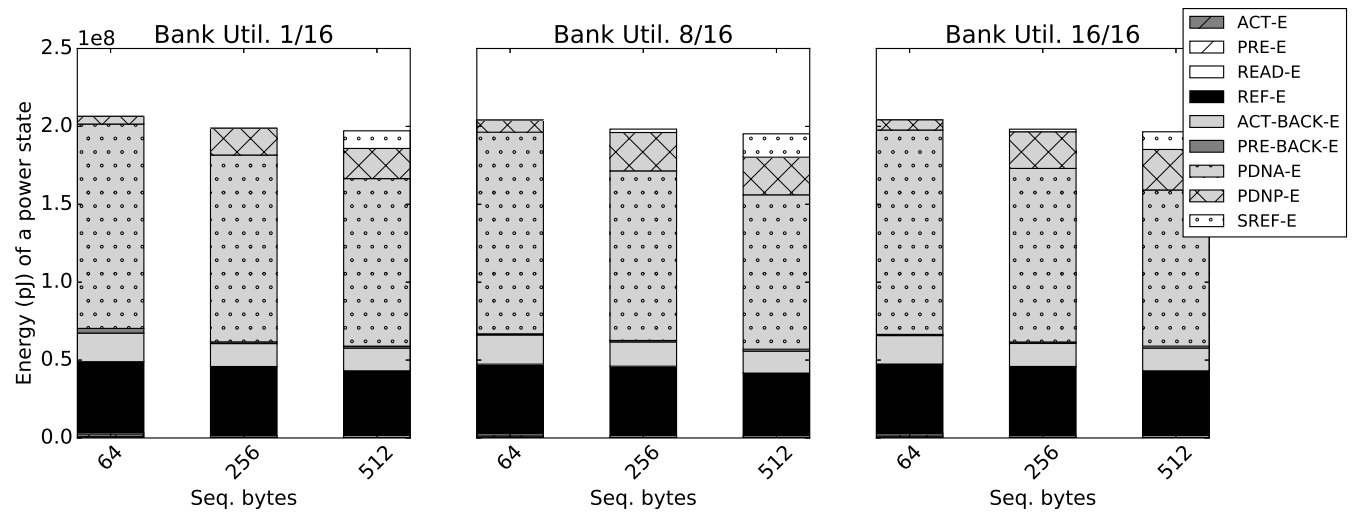

(a) Open adaptive policy, dense traffic

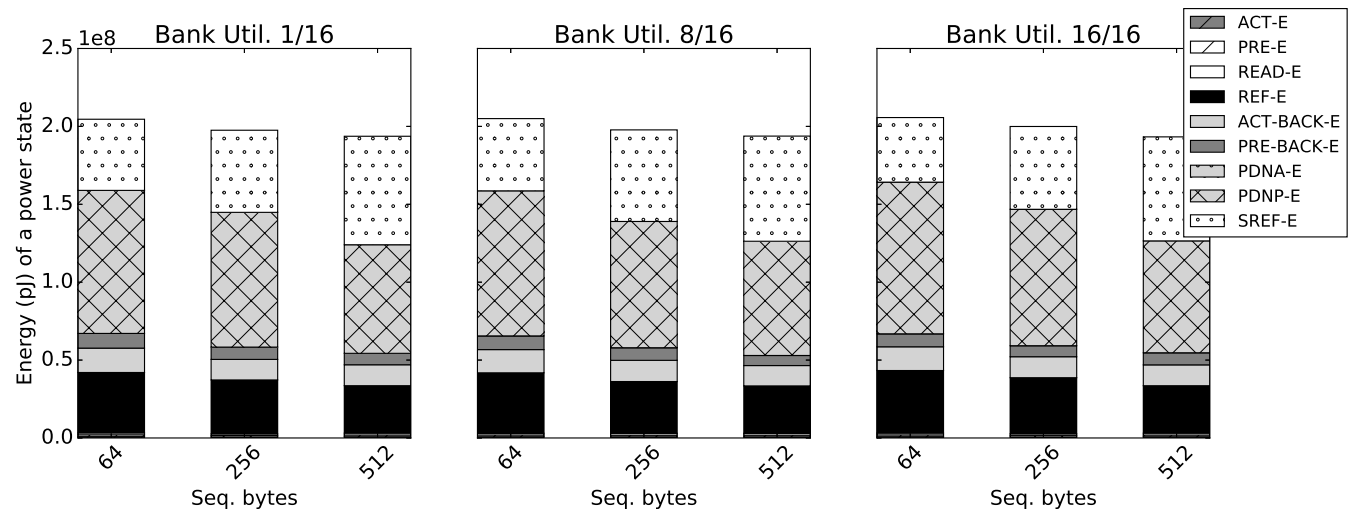

(b) Closed adaptive policy, dense traffic

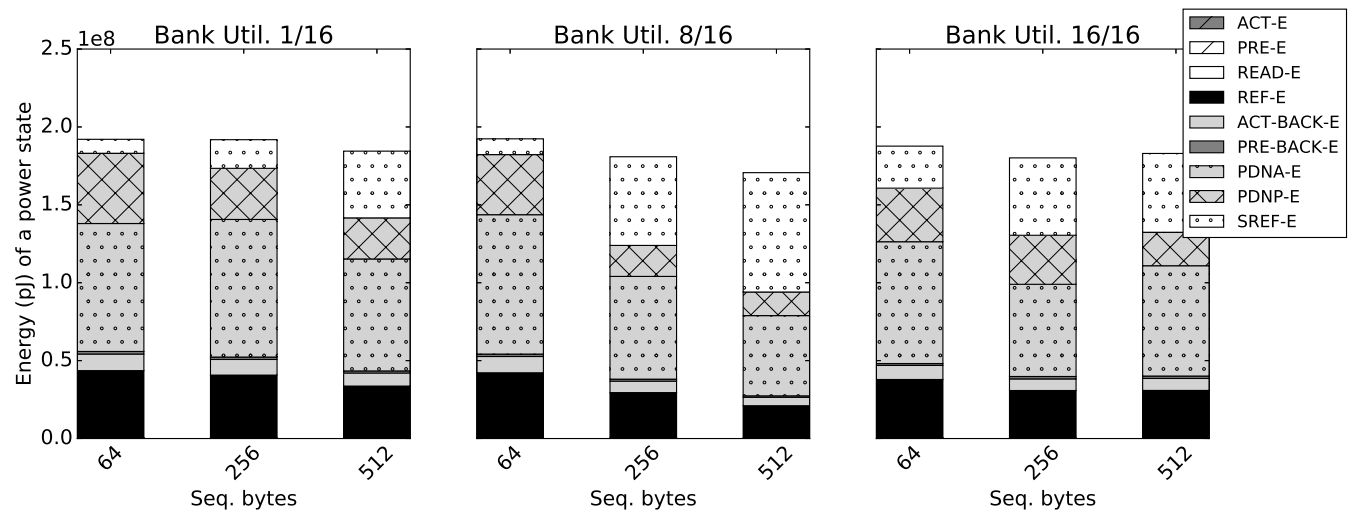

(c) Open adaptive policy, sparse traffic

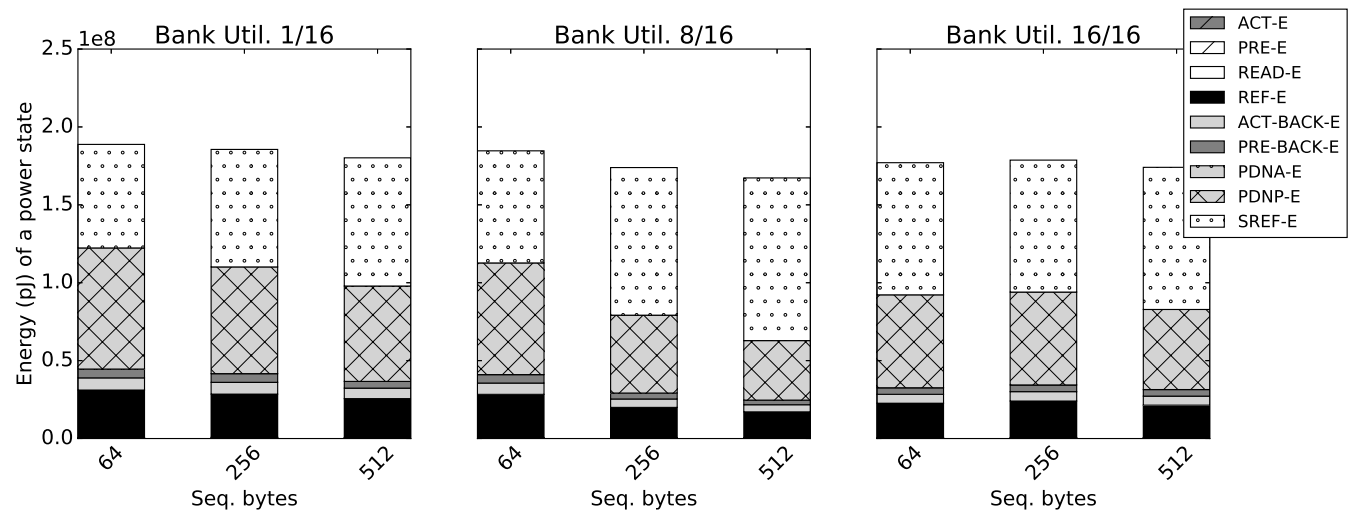

(d) Closed adaptive policy, sparse traffic

**Figure 6: Energy consumed by power states for $N_{Ranks}$ = 2**

the closed adaptive page policy and dense traffic (Figure 5b), the time spent in ACT is constant as $N_{SeqBytes}$ increases because if there are no queued requests the page is closed and therefore for each new request the bank needs to be activated. On the other hand, for the same traffic profile and open page policy (Figure 5a) we see a slight decrease in the activate time because if there are no queued requests the page is kept open and there are row hits when $N_{SeqBytes}$ is 256 and 512.

Thus by tuning sparseness of requests, we validate the model regarding time spent in different power states by analyzing across row hit behaviour, bank utilization and page policies. The corresponding energy components are also in line with the current values used.

# 5 EVALUATION WITH HPC APPLICATIONS

Having validated the behaviour of the new power-down functionality added to the DRAM controller in gem5, we evaluate the model for realistic workload.

## 5.1 Applications

We use the following proxy applications from the High Performance Compute (HPC) domain.

(1) **HPCG** stands for High Performance Conjugate Gradient and represents computational and data access patterns that match a broad set of HPC applications, including sparse matrix-vector multiplication and Gauss-Seidel smoother among others [10].
(2) **FFT** measures the floating point rate of execution of double precision complex one-dimensional Discrete Fourier Transform. [43]
(3) **Pathfinder** is a signature search algorithm in which is a modified depth-first recursive search wherein adjacent nodes are compared before recursing down its edges for labels [9].
(4) **GUPS** stands for Giga Updates Per Second and measures the rate of integer random updates of memory [31].
(5) **DGEMM** measures the floating point rate of execution of double precision real matrix-matrix multiplication [31].
(6) **Linpack** consists of an algorithm that solves a (random) dense linear system in double precision arithmetic [38].

## 5.2 Experimental Setup

We use the *Elastic Traces* methodology which models traffic from an out-of-order processor with reasonable timing accuracy. The advantage of elastic trace replay is faster simulation speed compared to typical execution-based cycle-level processor model [19]. We generate single-core traces for the region of interest for the aforementioned applications. FFT and GUPS are traced for 200 and 100 million instructions respectively. All other applications are traced for 1 billion instructions.

The DRAM memory sub-system configuration is 2 ranks and open adaptive page policy with the same timing and power parameters as in Section 4. The un-core part of the system includes a two-level cache hierarchy and a crossbar as the system interconnect. We perform two simulations - one with the power-down functionality we added to the DRAM controller model and one without. By comparing the execution time and energy consumption for the two models we can analyze the trade-offs.

## 5.3 Results

| | % of total energy | | | | | |
|---|---|---|---|---|---|---|
| | HPCG | FFT | Pathfinder | GUPS | DGEMM | Linpack |
| PDNA | 42% | 28% | 51% | 38% | 27% | 27% |
| PDNP | 3% | 8% | 3% | 0% | 5% | 7% |
| SREF | 4% | 14% | 0% | 0% | 28% | 23% |
| Total | 49% | 50% | 54% | 38% | 60% | 57% |

**Table 3: Energy in low power modes PDNA, PDNP and SREF as a percentage of total energy across all applications.**

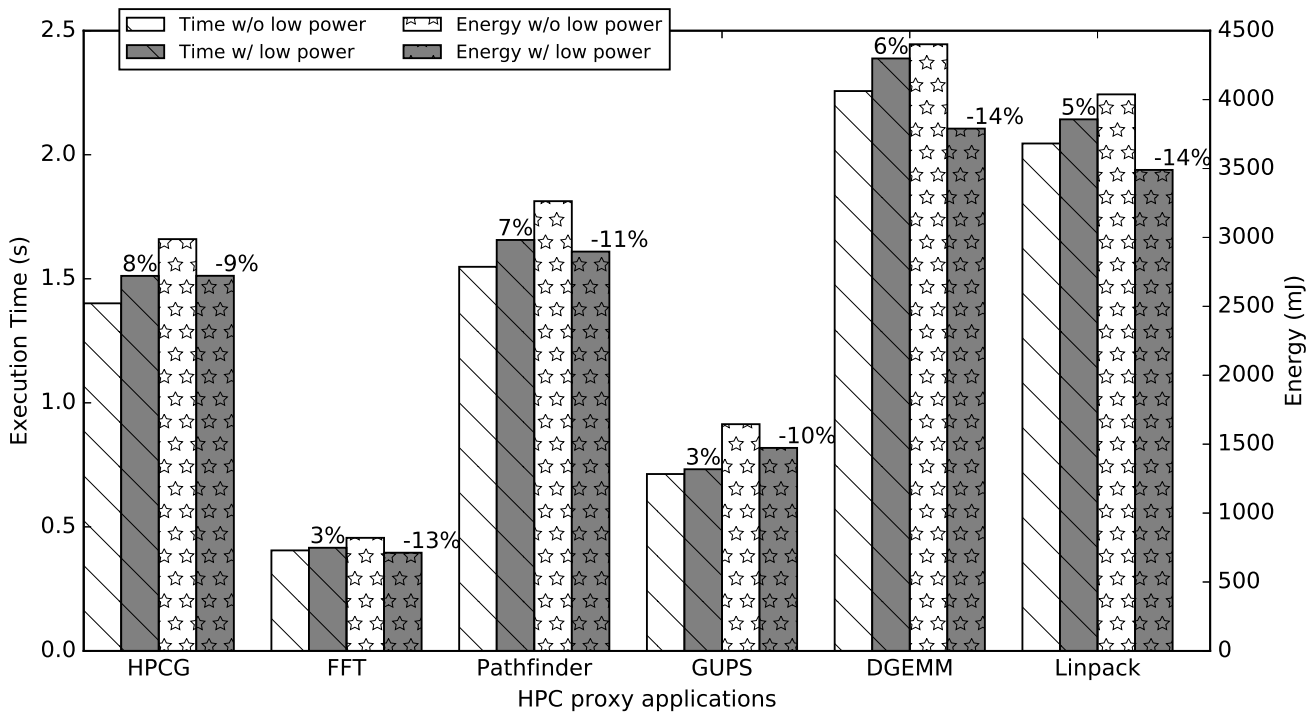

**Figure 7: Execution time and energy of DDR4-2400 with and without power-down functionality.**

Figure 7 shows the execution time and total energy consumed with and without the power-down functionality in the DRAM controller. The labels on the grey bars represent the percentage difference. The energy consumption reduces by 9%-14% while the execution time increases by 3%-8%. The impact of power-down modes on energy is in line with the staggered power-down results [26].

Table 3 shows the percentage of energy consumed in low power modes PDNA, PDNP and SREF with respect to the total energy for all applications. The applications FFT, DGEMM and Linpack are compute intensive with well behaved memory access patterns which means long periods of no DRAM activity. Therefore, for FFT, DGEMM and Linpack, we see higher percentages of energy consumed by low power modes (50%, 60% and 57% respectively). In Figure 7, these applications show the least execution time increase due to power-down modes (3%, 6% and 5%) and the maximum savings in energy (13%, 14% and 14%).

HPCG and Pathfinder have a bursty memory behaviour and see very little savings coming from PDNP and SREF. Finally, GUPS is a synthetic benchmark designed to stress the DRAM with random memory accesses. As expected, there is no energy consumed by PDNP and SREF and 38% energy consumed by PDNA.

While an HPC system would have many cores, using one core lets us better correlate the access patterns of the workload to expected low power behavior. Future work includes analysis on larger systems for HPC as well as extending to other applications like mobile and server workloads. The above results illustrate how the newly added low power functionality can enable the research community to perform DRAM energy efficiency studies for a breadth of systems using gem5 - an open-source and publicly available simulator.

# 6 RELATED WORK

We now cover the related work in this area of research, which falls into two categories.

## 6.1 DRAM Power-Down

In existing power-down techniques for DDR DRAMs, the memory controller monitors the current memory traffic and predicts when to power down and into which power-down state. Due to a limited buffer depth, the memory controller has only a *local* view on the executed application [22]. Therefore, the full advantages of the power-down states implemented in modern DRAMs cannot be exploited efficiently because the memory controller lacks information with respect to the application's memory access behavior. Moreover, the idle periods between two consecutive memory accesses are often too short to exploit, e.g., the full potential of self-refresh [15]. Therefore, memory controllers usually issue power-down commands after configurable timeouts [12, 30]. Thus, the authors of [15] propose a technique that actively reshapes the memory traffic in order to merge short idle periods, which were previously too short for effective power-down management. Like this, they can effectively exploit the idleness in the DRAM. Fan et al. [12] discovered that the DRAM should be immediately transitioning to a lower power state when it becomes idle.

In a comprehensive approach, Hur et al. [16] use the memory scheduler to improve the usage of the power-down modes and to throttle DRAM activity based on predicted delays caused by the throttling while still keeping the performance sufficiently high. In [2] memory pages are loaded only to few memory DIMMs, while the the rest of the DIMMs are switched to self-refresh. In [8], compiler techniques are presented, which exploit the DRAM power-down states. Chandrasekar et al. [5] present two power-down strategies that reduce memory energy consumption while still preserving the guaranteed bandwidth provided by real-time memory controllers. Furthermore, they present an algorithm to select the most energy-efficient power-down mode at run-time. In [44], a history-based predictor is used in order to forecast the duration of an idle period. According to the forecast, either the self-refresh mode, normal power-down or a combination of both is selected. However, such a predictor implies significant hardware overhead in the front-end of the memory controller. The authors of [32] present a post-DDR4 DRAM architecture that is capable of fast wake-up while ensuring high bandwidth. Jung et al. present a technique called *Staggered Power-Down* [26]. This approach does neither require changes to the DRAM architecture, nor sophisticated predictors in the memory controller. Their approach uses the internal refresh timer for the transition of the power-down states. Thus, dedicated hardware timeout counters, as used in state-of-the-art controllers, are not required anymore. Thus this work uses the *Staggered Power-Down* approach.

## 6.2 DRAM- and Power-Down Simulation

Many MPSoC simulators use detailed cache and interconnect models, but assume a simplistic *Fixed-Latency* memory model [17]. In the fixed latency model, all memory requests experience the same latency. Queueing, scheduling, or reordering of memory requests in the memory controller is not modelled. Thus, always the maximum memory bandwidth is considered, which is completely unrealistic since the latency of a DRAM access varies between a dozen and several hundred cycles.

When it comes to high-level simulations of DRAM subsystems, several DRAM simulators exists [7, 20, 28, 40]. Each of them is focusing on different aspects like, e.g., scheduling, subsystem architecture etc., and each has individual advantages and drawbacks. They either do not implement a power-down functionality, do not implement it truthfully or completely, or if so they are not maintained anymore with respect to today's DRAM standards. Moreover, they all are cycle accurate simulators that slow down event-driven full-system simulations, because every cycle has to be simulated. As mentioned before, since DRAMs contribute significantly to the power consumption of today's systems, there is a need for accurate power modelling. A well-known and often used DRAM power model is provided by Micron in form of a spreadsheet [37]. A more accurate model is DRAMPower, by Chandrasekar et al. [6], which uses the actual timings from stimuli generated by a functional DRAM simulator. Moreover, DRAMPower provides the possibility to estimate the power consumption of the DRAM power-down modes.

Jung et al. presented DRAMSys [24, 25], a SystemC/TLM based DRAM subsystem design space exploration framework featuring functional, power and thermal modeling (based on DRAMPower and 3D-ICE [42]), a detailed power-down model [26], as well as a sophisticated retention error model [45]. However, at present this tool is not publicly available.

*gem5* [3], a full-system simulator has integrated a realistic DRAM controller model [14]. This is very similar to the one implemented in DRAMsys and also uses the DRAMPower library [6]. Furthermore, *gem5* features elastic traces [19] which enable fast and reasonably accurate memory subsystem design space exploration.

Prior to our work the integration of DRAM power-down modes was missing in *gem5*. However, as part of publishing this paper we contributed the DRAM power-down functionality to the *gem5* simulator. Due to our code contribution, and the validation shown in Sections 4 and 5, we provide the research community with a publicly available full-system simulator for energy efficiency studies.

# 7 CONCLUSION

DRAM is a significant contributor to the total energy in systems from smartphones to servers. DRAM technology incorporates power-down modes which enable architects to design power-down strategies when there is no memory activity and thus improve the energy efficiency of these systems. From existing publicly available simulators, there is none that includes a fully functional DRAM sub-system, a sophisticated controller, power-down modes, an integrated power calculation tool and is able simulate complete SoCs with real workloads (see Section 6).

In the context of this paper, we describe the power-down functionality that we contributed to the publicly available and open source full-system simulator gem5 [3], which already had a controller and integrated power calculation tool. We describe the power state machine (see Section 3) and validate the newly added functionality systematically by using synthetic traffic designed to put the device in power-down modes for varying amounts of time. We analyze the behaviours across open adaptive and closed adaptive page policies as well as for increasing bank utilization and row hits (see Section 4). In addition, we evaluate the model using real workloads belonging to the HPC domain (see Section 5) to show that the value of our contribution is in enabling the research community to perform energy efficiency studies for a range of problems using a full-system simulator.

## ACKNOWLEDGMENTS

This work was partially funded by the German Research Foundation (DFG) grant no. WE2442/10-1 (http://www.uni-kl.de/3d-dram) and supported by the the *Fraunhofer High Performance Center for Simulation- and Software-based Innovation*. The project OPRECOMP acknowledges the financial support of the Future and Emerging Technologies (FET) programme within the European Unions Horizon 2020 research and innovation programme, under grant agreement No.732631 (http://www.oprecomp.eu)

The authors thank Omar Naji for the initial basic model of power-down modes. The authors are also grateful to Stephan Diestelhorst, Andreas Sandberg and Nikos Nikoleris for their valuable feedback on the writing and the code contributed to gem5.